\pdfoutput=1
\documentclass[aps,pra,twocolumn,10pt]{revtex4-1}
\usepackage[utf8]{inputenc}
\usepackage{cochineal}
\usepackage[cochineal]{newtxmath}
\usepackage{mathtools,graphicx,microtype,siunitx}
\usepackage[cal=pxtx]{mathalfa}
\usepackage[colorlinks=true,allcolors=blue]{hyperref}

\begin{document}

\title{Spin-orbit-driven electron pairing in two dimensions}

\author{Yasha Gindikin and Vladimir A.\ Sablikov}

\affiliation{Kotel'nikov Institute of Radio Engineering and Electronics,
Russian Academy of Sciences, Fryazino, Moscow District, 141190, Russia}

\begin{abstract}
We show that the spin-orbit interaction (SOI) arising due to the in-plane electric field of the Coulomb repulsion between electrons in a two-dimensional quantum well produces an attractive component in the pair interaction Hamiltonian that depends on the spins and momenta of electrons. If the Rashba SOI constant of the material is high enough the attractive component overcomes the Coulomb repulsion and the centrifugal barrier, which leads to the formation of the two-electron bound states. There are two distinct types of two-electron bound states. The \textit{relative} bound states are formed by the electrons orbiting around their common barycenter. They have the triplet spin structure and are independent of the center-of-mass momentum. In contrast, the \textit{convective} bound states are formed because of the center-of-mass motion, which couples the electrons with opposite spins. The binding energy in the meV range is attainable for realistic conditions.
\end{abstract}

\maketitle

\section{Introduction}
The fact that the spin-orbit interaction (SOI) is produced by Coulomb fields of interacting electrons was established by Breit in the context of a two-particle problem in relativistic quantum mechanics~\cite{bethe2012quantum}. However, this effect is very small on the scale of the electron energy in solids. The SOI created by the electric fields in crystals is known to be much stronger than in vacuum~\cite{winkler}. Therefore, the SOI produced by the Coulomb field of interacting electrons can also be strong enough to give rise to new nontrivial effects. The SOI effect of this origin was first demonstrated by McLaughlan \textit{et al.}~\cite{0953-8984-16-39-017}. They showed that the interaction of electrons with the image charges leads to the Rashba-like SOI splitting of the electronic states near the surface of metals.

Recently, we have shown that in low-dimensional systems the SOI produced by the Coulomb interaction of electrons with the image charges induced on a nearby metallic gate leads to nontrivial effects in materials with large Rashba SOI constants~\cite{PhysRevB.95.045138,doi:10.1002/pssr.201700256,2017arXiv170700316G}. The main effect is that because of this interaction, the electron-electron (\textit{e-e}) interaction Hamiltonian acquires a spin-dependent component that is attractive for a particular spin orientation locked to momentum. As a result of such attraction, one of the collective modes of a many-electron system is softened, and a homogeneous ground state becomes unstable as the Rashba SOI constant exceeds a critical value~\cite{PhysRevB.95.045138}.

A remarkable feature of this mechanism of the electron attraction is that it is determined by a combined effect of the motion of electrons and their Coulomb interaction so that the pair interaction depends on the electron spins and momenta. Under definite conditions this attraction can lead to the pairing of electrons. Given the spin-selective nature of the attractive SOI, a number of bound states with different spin structure can be expected. 

We have studied this pairing mechanism by solving a two-body problem for a one-dimensional electron system to show that there exist two distinct types of bound states of electrons, depending on the type of their motion~\cite{2018arXiv180410826G}. Due to the relative motion, the bound states are formed by electrons with opposite spins, whereas the motion of the electron pair as a whole creates bound states with parallel spins. 

In a two-dimensional (2D) system, the situation is more complicated since in addition to a normal electric field there exists an in-plane electric field that also produces the Rashba SOI\@. The possibility of electron pair formation in a 2D gated system was considered in Ref.~\cite{chaplik}, disregarding the in-plane Coulomb field. In this case, only one type of bound state was discussed. The effect of the in-plane electric field has not been studied yet, but it is clear that it can also be essential, and the relative role played by both components of the Coulomb field depends on the parameters of the system. 

In the present work, we consider the problem of two interacting electrons in a 2D system with the SOI arising solely from the in-plane Coulomb field. We show that the spin-dependent attractive component emerges in the \textit{e-e} interaction Hamiltonian due to the SOI without any intervention of the image charges. The two-electron bound states can appear if the SOI overcomes the Coulomb repulsion. This happens for sufficiently thin 2D layers hosting the electrons or for the large enough values of the Rashba parameter. Under these conditions, there exist two distinct types of two-electron bound states in which the electrons are moving differently. 

The \textit{relative bound states} are triplet pairs formed by the electrons orbiting around their common barycenter, with a spin orientation locked to the orbital angular momentum. The binding energy of these states does not depend on the motion of the center of mass.

The \textit{convective bound states} appear because of the motion of the electron pair as a whole, which couples the electrons with opposite spins. The effective attraction is growing with the center-of-mass momentum so that the binding energy and the effective mass of the pair essentially depend on its momentum.

\section{The model}
Consider two electrons in a 2D quantum well situated in the $x$-$y$ plane.The kinetic energy is 
\begin{equation}
\label{kin}
	H_{\mathrm{kin}} =\sum_{i=1}^{2} \frac{\mathbf{p}_i^2}{2m}\,,
\end{equation}
with $\mathbf{p}_i= -i \hbar\nabla_{\mathbf{r}_i}$ being the momentum, $\mathbf{r}_i = (x_i,y_i)$ being the position of the $i$th electron, and $m$ is the effective electron mass. The \textit{e-e} interaction Hamiltonian consists of two components. The first part is the Coulomb \textit{e-e} repulsion described by the interaction potential,
\begin{equation}
\label{Coul}
	U(\mathbf{r}_1 - \mathbf{r}_2) = \frac{e^2}{\epsilon |\mathbf{r}_1 - \mathbf{r}_2|}\,,
\end{equation}
$\epsilon$ being the dielectric constant. The second part is the SOI,
\begin{equation}
\label{SOI}
	H_{\mathrm{SOI}} = \frac{\alpha}{\hbar} \sum_{i \ne j}  \left[E_y(\mathbf{r}_i - \mathbf{r}_j) p_{i x} - E_x(\mathbf{r}_i - \mathbf{r}_j) p_{i y} \right] \sigma_{z_i}\,,
\end{equation}
produced by the electric field 
\begin{equation}
\label{fld}
	\mathbf{E}(\mathbf{r}_i - \mathbf{r}_j) = -e \frac{\mathbf{r}_i - \mathbf{r}_j}{\epsilon |\mathbf{r}_i - \mathbf{r}_j|^3}
\end{equation}
acting on the $i$th electron from the $j$th electron. Here $\sigma_{z_i}$ is the Pauli matrix and $\alpha$ is the SOI constant.

It is important that the SOI of Eq.~\eqref{SOI} is a two-particle interaction. The sign of this interaction depends on the product of electron momentum and spin projections, thus the interaction is attractive for a particular spin orientation tied to momentum.

The two-electron wave function is a Pauli spinor of the fourth rank, $\Psi(\mathbf{r}_1,\mathbf{r}_2) = {\left(\Psi_{\uparrow \uparrow},\Psi_{\uparrow \downarrow},\Psi_{\downarrow \uparrow},\Psi_{\downarrow \downarrow}\right)}^{\intercal}$. The full Hamiltonian built as a Kronecker  sum from Eqs.~\eqref{kin}--\eqref{fld} is diagonal in the corresponding basis, so the Schr\"odinger equation for $\Psi(\mathbf{r}_1,\mathbf{r}_2)$ splits into four separate equations for the spinor components. 

Let us switch from the positions of the individual electrons to the relative position $\mathbf{r} = \mathbf{r}_1 - \mathbf{r}_2$ and the center-of-mass position $\mathbf{R} = (\mathbf{r}_1 + \mathbf{r}_2)/2$. Denote the corresponding momentum operators by $\mathbf{p}$ and $\mathbf{P}$, respectively.

Then the equations for $\Psi_{\uparrow \uparrow}$ and $\Psi_{\uparrow \downarrow}$ are
\begin{equation}
\label{rel}
	\begin{split}
		\left[- \frac{\hbar^2}{m} \Delta_{\mathbf{r}} - \frac{\hbar^2}{4m} \Delta_{\mathbf{R}} + U(\mathbf{r}) + \frac{2\alpha}{\hbar} {(\mathbf{r} \times \mathbf {p})}_z \frac{e}{\epsilon r^3} \right] &\Psi_{\uparrow \uparrow}\\ 
		= \varepsilon_{\uparrow \uparrow}  &\Psi_{\uparrow \uparrow}
	\end{split}
\end{equation}
and
\begin{equation}
\label{conv}
	\begin{split}
		\left[- \frac{\hbar^2}{m} \Delta_{\mathbf{r}} - \frac{\hbar^2}{4m} \Delta_{\mathbf{R}} + U(\mathbf{r}) + \frac{\alpha}{\hbar} {(\mathbf{r} \times \mathbf {P})}_z \frac{e}{\epsilon r^3} \right] &\Psi_{\uparrow \downarrow}\\
		= \varepsilon_{\uparrow \downarrow}  &\Psi_{\uparrow \downarrow}\,.
	\end{split}
\end{equation}
The equations for $\Psi_{\downarrow \downarrow}$ and $\Psi_{\downarrow \uparrow}$ can be obtained from the above equations by changing the sign of $\alpha$. The solution of the system should be antisymmetrized with respect to the particle permutation.

It is clear that in Eq.~\eqref{rel} the relative motion of electrons is separated from the motion of their center of mass. Therefore the solutions of this equation are the relative states. The SOI term in Eq.~\eqref{rel} is similar to the SOI in atoms~\cite{landau1958course} except for the difference in the magnitude of $\alpha$ and the dimensionality of the system. 

On the contrary, in Eq.~\eqref{conv} the relative motion is not separated from the motion of the center of mass. Moreover, it is the center-of-mass motion that determines the SOI term which describes the attraction between the electrons for a particular spin configuration and finally leads to the formation of bound states. These are the convective bound states. Both kinds of the bound states are investigated in detail below.

\section{Relative bound states}
The center-of-mass motion fully decouples from Eq.~\eqref{rel}, so that $\Psi_{\uparrow \uparrow}(\mathbf{r},\mathbf{R}) = \psi_{\uparrow \uparrow}(\mathbf{r}) \exp(i \mathbf{K} \cdot \mathbf{R})$, with the wave function of the relative motion satisfying
\begin{equation}
\label{rel1}
		\left[- \frac{\hbar^2}{m} \Delta_{\mathbf{r}} + U(\mathbf{r}) + 2 \alpha l_z \frac{e}{\epsilon r^3} \right]\psi_{\uparrow \uparrow}(\mathbf{r}) = \varepsilon \psi_{\uparrow \uparrow}(\mathbf{r})\,.
\end{equation}
Since the orbital angular momentum along the $z$ direction $l_z = - i \partial_{\phi}$ commutes with the Hamiltonian, the solution of Eq.~\eqref{rel1} can be chosen as the eigenfunction of $l_z$,
\begin{equation}
\label{simplel}
	\psi_{\uparrow \uparrow}(\mathbf{r}) = \frac{u(r)}{\sqrt{r}} e^{i l \phi}\,. 
\end{equation}
The orbital angular quantum number $l$ should be an odd integer because of the antisymmetric properties of $\Psi_{\uparrow \uparrow}$, which should change sign upon the electron permutation, i.e., as $\phi \to \phi + \pi$. The even values of $l$ are not allowed, therefore, e.g.,\ the $s$ states do not exist, in contrast to the 2D hydrogen atom~\cite{PhysRevA.43.1186}. Most importantly, the SOI lifts the Coulomb degeneracy in $l$.

Let us normalize the distance to the Bohr radius $a_B = \epsilon \hbar^2/me^2$, the energy to the  Rydberg constant in the material $Ry = \hbar^2/2 m a_B^2$, and let us introduce the dimensionless SOI constant $\tilde{\alpha} = \alpha/e a_B^2$.

Then the equation for the radial part $u(r)$ takes the form
\begin{equation}
\label{releq}
	-\frac{\partial^2 u}{\partial r^2} + V(r) u = \frac{\varepsilon}{2} u\,,
\end{equation}
where the binding potential is
\begin{equation}
\label{relsing}
	V(r) = \frac{l^2 - \frac{1}{4}}{r^2} + \frac{1}{r} + \frac{2 \tilde{\alpha} l}{r^3}\,.
\end{equation}
The first term of the binding potential is a repulsive centrifugal potential, while the second comes from the repulsive Coulomb interaction. The third term is produced by the SOI and can be either repulsive or attractive depending on the sign of $l$. Negative $l$ supports the existence of the relative bound state $\Psi_{\uparrow \uparrow}$, whereas positive $l$ supports $\Psi_{\downarrow \downarrow}$. Thus the spin orientation of this triplet state is locked to the orbital angular momentum.

The attractive $r^{-3}$ singularity in the potential of Eq.~\eqref{relsing} should be regularized to avoid the ``fall to the center''~\cite{landau1958course}. At small distances between the electrons the Eqs.~\eqref{Coul} and~\eqref{fld} should be modified for two reasons. First, to account for the finite layer thickness $d$, the Coulomb interaction potential is approximated as
\begin{equation}
\label{Coulreg}
	U(\mathbf{r}) = \frac{e^2}{\epsilon \sqrt{r^2 + d^2}}
\end{equation}
and the field as
\begin{equation}
\label{Fieldreg}
	\mathbf{E}(\mathbf{r}) = \frac{1}{e}\nabla U(\mathbf{r}) = - e\frac{\mathbf{r}}{\epsilon {\left(r^2 + d^2\right)}^{\frac{3}{2}}}\,.
\end{equation}
A second mechanism of cutting off the potential and the SOI magnitude at small distances is related to the Zitterbewegung of electrons in crystalline solids~\cite{0953-8984-23-14-143201}, which we take into account phenomenologically similar to Eqs.~\eqref{Coulreg} and~\eqref{Fieldreg}. In what follows, we treat $d$ as a combined short-distance cutoff.

Thus the regularized binding potential becomes (in normalized units)
\begin{equation}
\label{relreg}
	V(r) = \frac{l^2 - \frac{1}{4}}{r^2} + \frac{1}{\sqrt{r^2 + d^2}} + \frac{2 \tilde{\alpha} l}{{\left(r^2 + d^2\right)}^{\frac{3}{2}}}\,.
\end{equation}
The potential well is deepest for $|l| = 1$ as this minimizes the centrifugal barrier. Consequently, the triplet-like ground state wave functions for the relative states are given by 
\begin{equation}
	\Psi(\mathbf{r},\mathbf{R}) = {\left(e^{-i \phi},0,0,0\right)}^{\intercal}\frac{u(r)}{\sqrt{r}}e^{i \mathbf{K} \cdot \mathbf{R}}
\end{equation}
and
\begin{equation}
	\Psi(\mathbf{r},\mathbf{R}) = {\left(0,0,0,e^{i \phi}\right)}^{\intercal}\frac{u(r)}{\sqrt{r}}e^{i \mathbf{K} \cdot \mathbf{R}}\,.
\end{equation}

\begin{figure}[htb]
	\includegraphics[width=0.9\linewidth]{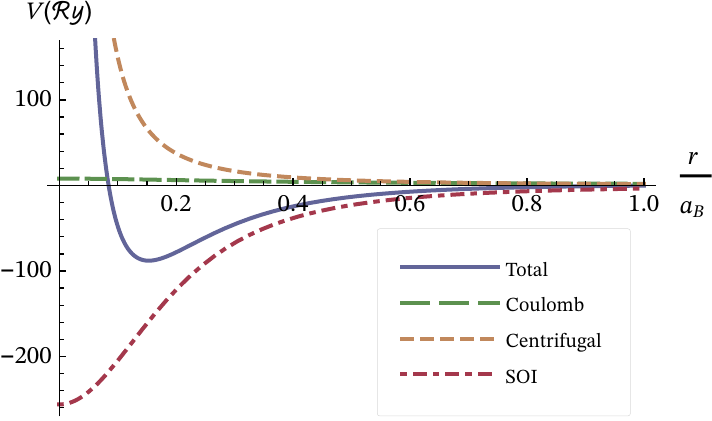}
		\caption{The effective binding potential for the relative bound states with $l = 1$ and separate contributions from the centrifugal potential, direct Coulomb \textit{e-e} interaction, and SOI\@.}
	\label{fig1}
\end{figure}
The potential profile is shown in Fig.~\ref{fig1} for $\tilde{\alpha} = 1$ and $d = 0.25 a_B$. The Coulomb repulsion is negligible compared to the centrifugal potential and SOI as long as $d \ll a_B$. In this limit, the binding potential is defined by a single parameter, $\alpha/d a_B$. Then the condition for the existence of the bound state is found to take a simple form: $\tilde{\alpha} > 2.3 \frac{d}{a_B}$. 

\begin{figure}[htb]
	\includegraphics[width=0.9\linewidth]{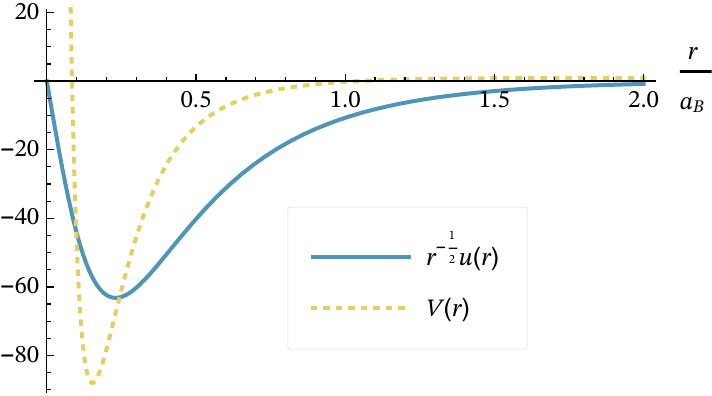}
		\caption{The radial part of the wave function of the relative bound state (up to the normalization constant) together with the effective binding potential for $d = 0.25 a_B$.}
	\label{fig2}
\end{figure}

Increasing the SOI parameter $\alpha$ or reducing the layer thickness $d$ increases the binding energy $|\varepsilon|$. For a layer thickness of $d = 0.25 a_B$, the energy of the relative bound state is $|\varepsilon| = \SI{4.5}{Ry}$ as calculated numerically from Eqs.~\eqref{releq} and~\eqref{relreg}. The size of the electron pair, estimated from the position of the peak in the radial wave function, is of the order of $d$. The spatial profile of the radial wave function together with the binding potential is shown in Fig.~\ref{fig2} for this case.

\section{Convective bound states}
The convective bound states appear as the solutions of Eq.~\eqref{conv}. The translational invariance implies that $\Psi_{\uparrow \downarrow}(\mathbf{r},\mathbf{R}) = \exp (i \mathbf{K} \cdot \mathbf{R}) \psi_{\uparrow \downarrow}(\mathbf{r},\mathbf{K})$, with $\mathbf{K}$ being a quantum number, but contrary to the previous case the center-of-mass motion essentially affects the relative motion. The wave-function of the relative motion $\psi_{\uparrow \downarrow}(\mathbf{r},\mathbf{K})$ depends on the center-of-mass wave vector $\mathbf{K}$ via the binding potential. 

We begin the study of the wave functions by considering the region of $r>d$ where the wave function behavior is of most interest. The analysis, the details of which are presented below in Sec.~\ref{RadialWF}, shows that in the region $r<d$ the wave function is extremely small. Therefore for $r>d$ we can use the uncut form of the potential. Using that form of the potential allows us to treat the problem analytically. The results of the numerical calculations with the smoothed form of the Coulomb potential and the electric field are presented in Sec.~\ref{Numerics}.

Let us direct the $y$ axis along $\mathbf{K}$ and denote the polar angle measured from the positive $x$-axis by $\phi$. The wave function is defined from the Schr\"odinger equation~\footnote{Note a similarity to the Schrödinger equation with the point-charge and dipole potentials. The pure dipole potential has quite a history; see K. Connolly and D. J. Griffiths, Am. J. Phys. \textbf{75}, 524 (2007), and references therein. The combined potential of the dipole and the point-charge was recently considered in M. Moumni and M. Falek, J. Math. Phys. \textbf{57}, 072104 (2016), in the exactly opposite case of attractive Coulomb interaction and weak dipole moment.},
\begin{equation}
\label{conveq}
	\begin{split}
		 \left[- \frac{1}{r} \frac{\partial}{\partial r}\left( r \frac{\partial}{\partial r} \right) -\frac{1}{r^2} \frac{\partial^2}{\partial \phi^2} + \frac{1}{r} +\frac{\mathcal{A} \cos \phi}{r^2} \right] & \psi_{\uparrow \downarrow}(\mathbf{r},\mathbf{K}) \\
		  = -\kappa^2 & \psi_{\uparrow \downarrow}(\mathbf{r},\mathbf{K})\,,
	\end{split}
\end{equation}
where $\kappa^2 = |\varepsilon|/2$ and the convenient dimensionless SOI constant is introduced,
\begin{equation}
	\mathcal{A} = \frac{\alpha K}{e a_B}\,.
\end{equation}
 
First two terms on the left hand side of Eq.~\eqref{conveq} are the kinetic energy with the centrifugal potential, and the third term is the Coulomb \textit{e-e} repulsion. We are mostly interested in the fourth term, which is exactly the SOI produced by the motion of the electron pair as a whole.

The SOI gives a strongly anisotropic contribution to the Hamiltonian, which does not commute with the orbital angular momentum. Hence $l$ is no longer a quantum number. This happens because the rotational symmetry is broken by the presence of the preferred direction along $\mathbf{K}$. As a result, the convective bound states acquire a non-trivial angular dependence different from Eq.~\eqref{simplel}.

Equation~\eqref{conveq} can be solved by the separation of variables,
\begin{equation}
	\psi_{\uparrow \downarrow}(\mathbf{r},\mathbf{K}) = f_{\uparrow \downarrow}(\phi) g(r)
\end{equation}
with the angular part given by
\begin{equation}
\label{mathieu}
	\frac{\partial^2 f_{\uparrow \downarrow}}{\partial \phi^2} + (-\lambda - \mathcal{A} \cos \phi)f_{\uparrow \downarrow} = 0
\end{equation}
and the radial part, same for $\psi_{\uparrow \downarrow}(\mathbf{r},\mathbf{K})$ and $\psi_{\downarrow \uparrow}(\mathbf{r},\mathbf{K})$, by
\begin{equation}
\label{radial}
	\left[- \frac{1}{r} \frac{\partial}{\partial r}\left( r \frac{\partial}{\partial r} \right) -\frac{\lambda}{r^2} + \frac{1}{r}\right]g =  -\kappa^2 g\,.
\end{equation}
The separation parameter $\lambda$ sets the binding potential magnitude. The sign and magnitude of $\lambda$ reflect the net effect of the attractive SOI competing with the repulsive centrifugal potential. Positive values of $\lambda$ correspond to the attractive binding potential for the radial motion.

\subsection{Angular dependence}

The $2 \pi$-periodic solutions of Eq.~\eqref{mathieu} arise for an infinite discrete set of values of $\lambda \in \{\, \lambda_m(\mathcal{A}) \mid m \in \mathbb{Z}\, \}$ that are related to the eigenvalues $a_{2m}(q)$ and $b_{2m}(q)$ corresponding to the Mathieu functions $ce_{2m}(z,q)$ and $se_{2m}(z,q)$~\cite{olver} via
\begin{equation}
	\lambda_m(\mathcal{A}) =
		\begin{dcases}
			-\frac{1}{4} a_{2m}(2\mathcal{A}), & m = 0,1,2 \ldots\,, \\
			-\frac{1}{4} b_{2|m|}(2\mathcal{A}), & m = -1,-2, \ldots \,.
		\end{dcases}
\end{equation}
The integer $m$ is an angular quantum number that supersedes the orbital quantum number $l$ for the convective states. The angular wave functions of the convective states are given by
\begin{equation}
	f_{\uparrow \downarrow}^{(m)}(\phi) =
		\begin{dcases}
			ce_{2m} \left(\frac{\phi}{2},2 \mathcal{A}\right), & m = 0,1,2 \ldots \,,\\
			se_{2 |m|} \left(\frac{\phi}{2},2 \mathcal{A}\right), &  m = -1,-2, \ldots\,.
		\end{dcases}
\end{equation}

\begin{figure}[htb]
	\includegraphics[width=0.9\linewidth]{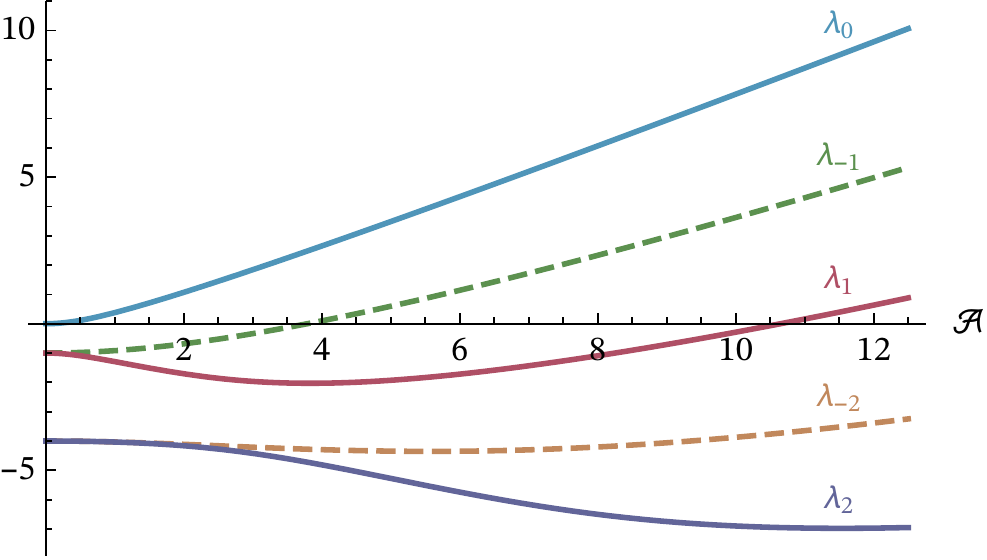}
		\caption{The dependence of $\lambda_m(\mathcal{A})$ on $\mathcal{A}$ for several values of $m$.}
	\label{fig3}
\end{figure}

Figure~\ref{fig3} shows the binding potential magnitude as a function of $\mathcal{A}$ for different quantum states. For a given $m$, $\lambda_m(\mathcal{A})$ is positive for sufficiently large $\mathcal{A}$, which means that the SOI of the large enough magnitude overcomes the repulsive centrifugal barrier to create an attractive potential for the radial motion. 

Note that $\lambda_0 > 0$ for any positive $\mathcal{A}$. However, because of the competing Coulomb repulsion, the bound states appear only for $\mathcal{A}$ exceeding some critical value, as determined in the next subsection. Still for a given $\mathcal{A}$ the potential well for the quantum state with $m=0$ is deeper than for $m \ne 0$. Thus the angular part of the ground state wave function $\psi_{\uparrow \downarrow}$ is 
\begin{equation}
\label{mathres}
	f_{\uparrow \downarrow}^{(0)}(\phi) = ce_0 \left(\frac{\phi}{2},2 \mathcal{A} \right)\,,
\end{equation}
with the corresponding binding potential magnitude being
\begin{equation}
\label{sep}
	\lambda_0(\mathcal{A}) = -\frac{1}{4} a_{0}(2 \mathcal{A})\,.
\end{equation}

\begin{figure}[htb]
	\includegraphics[width=0.9\linewidth]{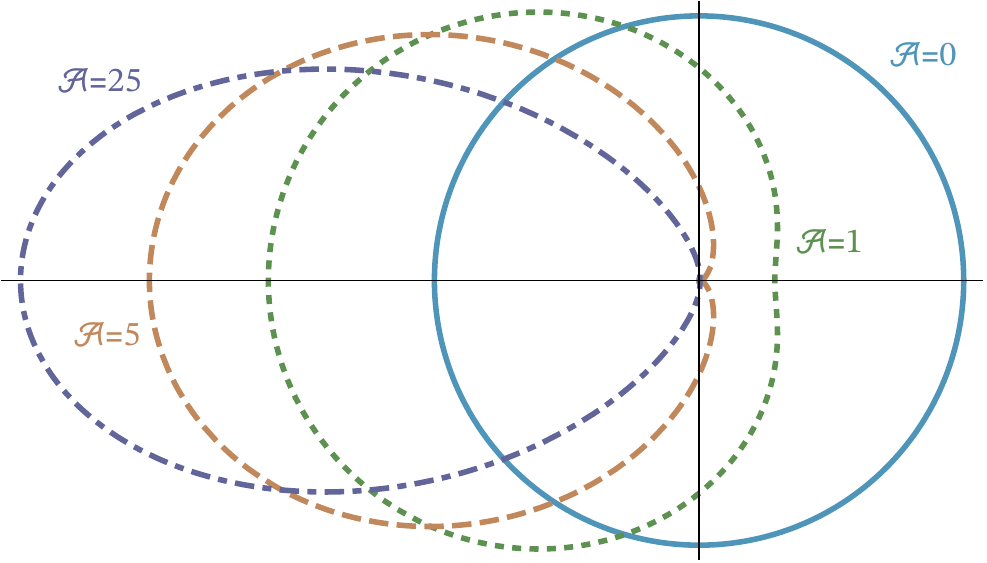}
		\caption{The polar diagram $\left(f_{\uparrow \downarrow}^{(0)}(\phi),\phi\right)$ for several values of $\mathcal{A}$.}
	\label{fig4}
\end{figure}

Figure~\ref{fig4} shows that as the SOI grows, $f_{\uparrow \downarrow}^{(0)}(\phi)$ evolves from a constant~\footnote{which formally corresponds to $l=0$, hence no centrifugal barrier $l^2/r^2$ to overcome} to a peak near $\phi = \pi$. The angular part of the $\psi_{\downarrow \uparrow}$ is
\begin{equation}
	f_{\downarrow \uparrow}^{(0)}(\phi) = ce_0 \left(\frac{\phi + \pi}{2},2 \mathcal{A} \right)
\end{equation}
with a peak near $\phi = 0$.

Note that $f_{\uparrow \downarrow}(\phi + \pi) = f_{\downarrow \uparrow}(\phi)$ and $f_{\downarrow \uparrow}(\phi + \pi) = f_{\uparrow \downarrow}(\phi)$. Consequently, the total antisymmetric wave function of the convective state is given by 
\begin{equation}
\label{convspinor}
	\Psi(\mathbf{r},\mathbf{R}) = {\left( 0,f_{\uparrow \downarrow}(\phi),-f_{\downarrow \uparrow}(\phi),0\right)}^{\intercal}g(r)e^{i \mathbf{K} \cdot \mathbf{R}}\,.
\end{equation}

\subsection{Radial dependence}
\label{RadialWF}

The attractive $-\lambda/r^{2}$ potential in Eq.~\eqref{radial} leads to the fall to the center~\cite{landau1958course}, unless properly regularized. A number of regularization techniques was developed~\cite{PhysRevLett.85.1590,PhysRevA.64.042103,PhysRevA.76.032112}, which are essentially based on introducing a short-distance cutoff~\cite{PhysRevD.48.5940}. 

We follow this approach by noting that in the region of $r<d$ the electric field of Eq.~\eqref{Fieldreg} linearly goes to zero with $r$, suppressing the attraction due to the SOI\@. In the same region, the Coulomb \textit{e-e} interaction potential of Eq.~\eqref{Coulreg} saturates at a finite positive value of $e^2/\epsilon d$, which gets large for small $d$. Consequently, a repulsive core is formed at $0< r <d$ by a combined action of the Coulomb repulsion and the centrifugal potential that reappears in the absence of SOI\@. On these grounds we regularize the potential $-\lambda/r^{2}$ by imposing a zero boundary condition for the radial wave-function,
\begin{equation}
\label{zerobound}
	g \big|_{r = d} =0\,,
\end{equation}
which defines the discrete spectrum of the convective states.

The solution of Eq.~\eqref{radial} is given by the Whittaker function~\cite{olver}. Up to the normalization constant, we have
\begin{equation}
\label{whitt}
	g(r) = r^{- \frac{1}{2}} W_{- \frac{1}{2 \kappa},i \sqrt{\lambda}} (2 \kappa r).
\end{equation}

According to the Sturm oscillation theorem~\cite{simon2005sturm}, the number of negative energy bound states is equal to the number of nodes of the zero-energy solution $g(r;\kappa = 0)$ in $(d, \infty)$. It is interesting that $g(r;\kappa = 0)$ belongs to the discrete part of the spectrum. The zero-energy bound state is protected by the cusp formed by the long-ranged Coulomb tail of the potential that approaches zero from the top as $r \to \infty$~\cite{DABOUL1994357}. Note that $g(r;\kappa = 0)$ can be expressed via the Macdonald function~\cite{olver}
\begin{equation}
	g(r;\kappa = 0) = \mathcal{K}_{2 i \sqrt{\lambda}}(2 \sqrt{r})\,.
\end{equation}
The wave function is normalizable since it behaves like $g(r;\kappa = 0) \sim r^{-\frac{1}{4}} \exp (- 2 \sqrt{r})$ as  $r \to \infty$. 

Use the boundary condition of Eq.~\eqref{zerobound} for $g(r;\kappa = 0)$ to define the critical magnitudes $\Lambda_n$ of the binding potential via
\begin{equation}
\label{ancrit}
	x_n \left(2 \sqrt{\Lambda_n}\right) = 2 \sqrt{d} \,,
\end{equation}
where $x_n (\mu)$ is the $n$th zero of the $\mathcal{K}_{i \mu}(x)$, $n = 1,2, \ldots$.
Then for a given angular quantum number $m$, the radial Eq.~\eqref{radial} has exactly $n$ bound states iff 
\begin{equation}
	\Lambda_n \leq \lambda_m(\mathcal{A}) < \Lambda_{n+1}\,.
\end{equation}
The states are indexed by a radial quantum number $k$ that takes a finite set of values, $k = 1 \ldots n$.

We obtain an analytical expression for the binding energy $\varepsilon_{k,m}$ by making a reasonable assumption that $d \ll a_B$, that is $d \ll 1$ in normalized units. Then for $\lambda \gtrsim 1$ the $1/r$ Coulomb repulsion is negligible compared to the attractive $-\lambda/r^{2}$ potential in Eq.~\eqref{radial} in the region of $d < r  \ll 1$. The wave function in this region is thus
\begin{equation}
	g(r) = \mathcal{K}_{i \sqrt{\lambda}}(\kappa r)\,.
\end{equation}
The spectrum of the convective states is defined from Eq.~\eqref{zerobound} to be
\begin{equation}
\label{anspect}
	|\varepsilon_{k,m}| = \frac{2 Ry}{d^2} x_k^2 \left(\sqrt{\lambda_m(\mathcal{A})} \right)\,.
\end{equation}
The ground state corresponds to $m = 0$, $k = 1$. An analytic approximation for $x_n(\mu)$ is given by Eq.~\eqref{Gosper}.

\subsection{Numerical results}
\label{Numerics}

Here we present the results of the direct numerical solution of the 2D Schr\"odinger equation~\eqref{conveq} with the smoothed form of the Coulomb \textit{e-e} interaction potential and electric field given by Eqs.~\eqref{Coulreg} and~\eqref{Fieldreg}. 

\begin{figure}[htb]
	\includegraphics[width=0.9\linewidth]{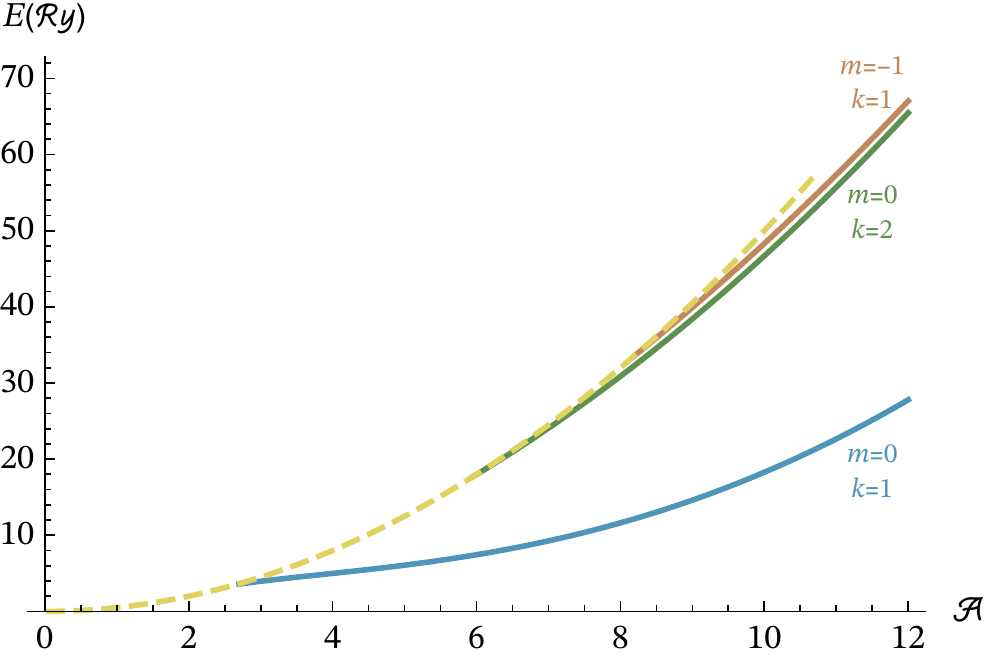}
		\caption{The system energy levels (solid lines) and the kinetic energy of the center of mass (dashed line) vs $\mathcal{A}$ for $d = 0.25 a_B$.}
	\label{fig5}
\end{figure}

Figure~\ref{fig5} shows the energies of the three convective states with different quantum numbers including the kinetic energy of the center of mass as a function of the parameter $\mathcal{A}$, which is proportional to both the Rashba SOI parameter and the center-of-mass momentum. Therefore, the lines in Fig.~\ref{fig5} also present the energy dispersion of the convective electron pair. The bound states appear in the spectrum at their respective critical values of $\mathcal{A}$. Their binding energy grows with $\mathcal{A}$.

Equation~\eqref{ancrit} leads to slightly larger critical values of $\mathcal{A}$, with qualitatively the same dependence of $E(\mathcal{A})$ given by Eq.~\eqref{anspect}.

\begin{figure}[htb]
	\includegraphics[width=0.9\linewidth]{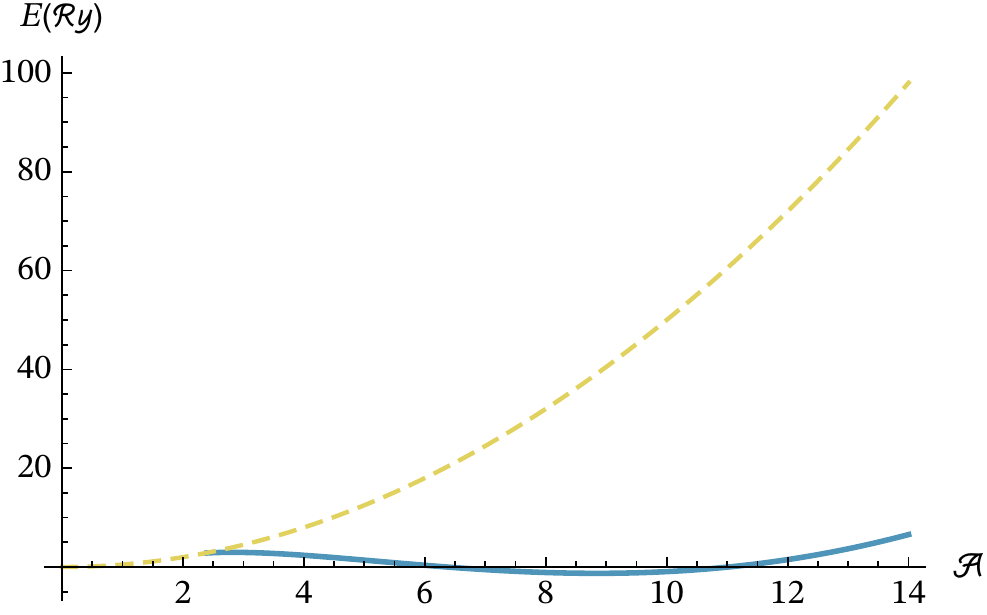}
		\caption{The ground state energy (solid line) and the kinetic energy of the center of mass (dashed line) vs $\mathcal{A}$ for $d = 0.2 a_B$.}
	\label{fig6}
\end{figure}

The effective mass of the electron pair is severely renormalized by the SOI and can even become negative as one lowers $d$, as can be seen from Fig.~\ref{fig6}.

\begin{figure}[htb]
	\includegraphics[width=0.9\linewidth]{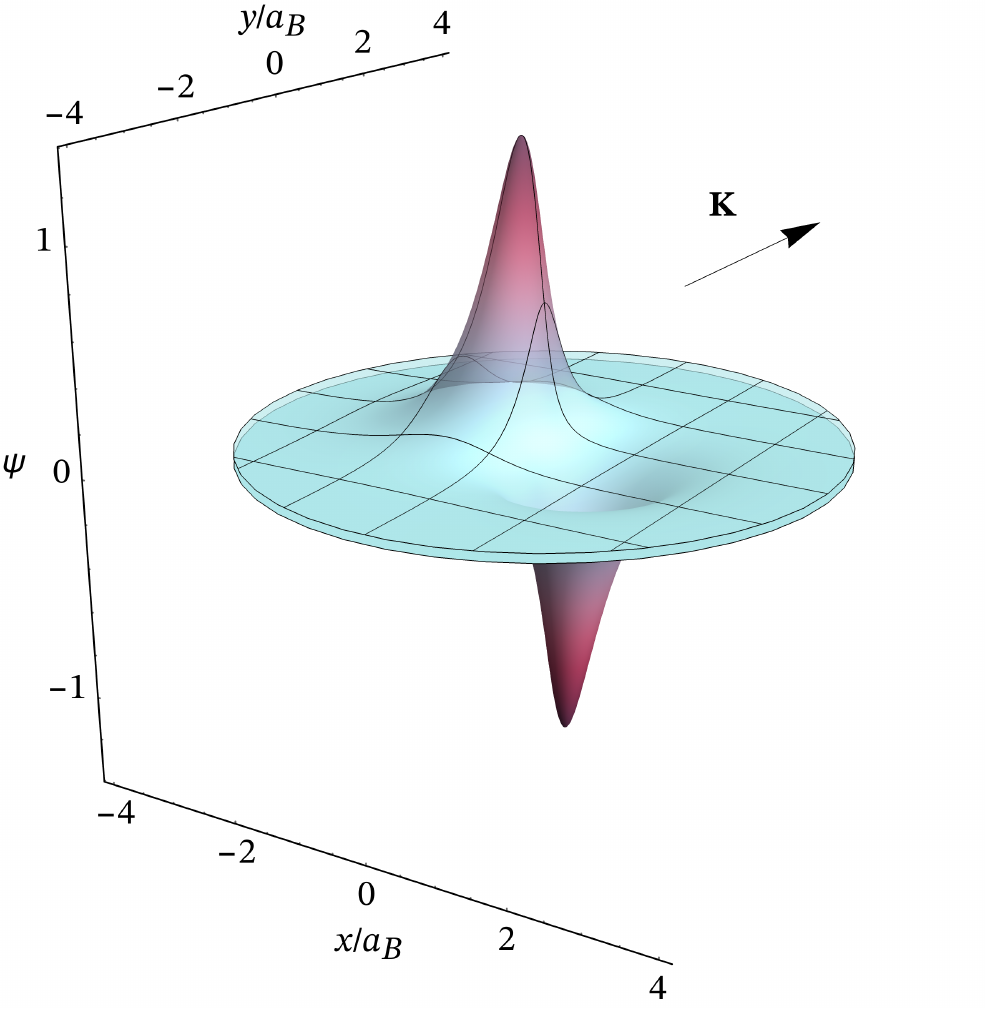}
		\caption{The spinor components of the convective state wave function for $d = 0.25 a_B$ and $\mathcal{A} = 5$ as functions of relative coordinates. Two surfaces are moved apart vertically by $\delta = \pm 0.02$ for better visual perception. The arrow shows the direction of vector $\mathbf{K}$ (along the $y$ axis).}
	\label{fig7}
\end{figure}

Figure~\ref{fig7} shows the wave function of the ground state, calculated numerically. Two surfaces combined in a single figure are the two spinor components $\psi_{\uparrow \downarrow}(\mathbf{r},\mathbf{K})$ and $\psi_{\downarrow \uparrow}(\mathbf{r},\mathbf{K})$. Note the strong dependence of the solution on the angle measured from the $\mathbf{K}$ direction, with peaks in the wave function shifted to the side off the line of motion. The analytic result of Eqs.~\eqref{mathres}--\eqref{whitt} leads to a similar picture.

\section{Conclusion}

We have proposed a new mechanism of electron pairing which stems from the spin-orbit component of the pair \textit{e-e} interaction. The effective attraction between electrons arises as a combined effect of the Coulomb field and the motion of electrons for certain configurations of their spins. This is principally different from the common mechanisms of electron pairing based on the renormalization of the \textit{e-e} interaction by many-particle excitations~\cite{kagan2013modern} or on the formation of a negative reduced effective mass of two electrons arising due to the peculiarities of the band spectrum~\cite{PhysRevB.95.085417}. The attraction of electrons stems from the well-known fact that the larger the electric field creating the SOI, the more strongly the SOI lowers the energy of electrons. As the distance between electrons shrinks, the electric field increases and therefore their energy lowers, which means that the electrons attract each other.

The fact that the bound state formation is determined by the electron momenta and spins leads to the highly unusual properties of the bound states. There exist two distinct types of bound states. 

The relative bound states depend only on the reciprocal electron motion. Their binding energy and electronic structure are unaffected by the motion of the electron pair as a whole. In 2D systems with in-plane Coulomb fields, the relative states are formed by electrons with parallel spins. 

The bound states of the other type are the convective states. In contrast to the relative states, they appear exactly because of the center-of-mass motion, which affects the relative motion of the electrons, the energy spectrum, and the spatial distribution of electron density. An astonishing property of these states is the non-trivial dependence of their energy on the momentum of the pair. As the momentum $K$ of the electron pair increases, the binding energy of the pair increases too, and it can become so large that the total energy of the pair starts to decrease with $K$. Thus, in some interval of $K$ the effective mass of the pair can become negative, which leads to dramatic consequences for the collective behavior of a many-electron system. This agrees qualitatively with a possible instability in a gated one-dimensional system~\cite{PhysRevB.95.045138}.

The existence conditions for the bound states impose rather serious requirements on the Rashba SOI constant $\alpha$ of the material and on the value of $d$ that defines the short-range cutoff of the binding potential, which are nonetheless attainable in currently available materials and conditions. Thus, taking $\tilde{\alpha} \approx 1$ and $a_B \approx \SI{100}{\angstrom}$, which is close to the parameters of, e.g., $\mathrm{Bi_2 Se_3}$~\cite{manchon2015new}, we obtain the binding energy of the relative state of the order of several meV for $d \approx \SI{30}{\angstrom}$. The localization scale and the binding energy to a substantial degree depend on $d$. A higher binding energy may be attainable for the states localized on a smaller spatial scale. However, to investigate this attractive possibility a different approach is required, which we are going to present in a follow-up work.

\acknowledgments
This work was partially supported by Russian Foundation for Basic Research (Grant No 17--02--00309) and Russian Academy of Sciences.

\appendix*
\section{}

The asymptotic formulas for the zeros $x_{n}$ of the Macdonald function $\mathcal{K}_{i \mu}(x)$ (also known as the modified Bessel function of the second kind) of pure imaginary order are known in the literature~\cite{FERREIRA2008223} for two limiting cases, $\mu \gg 1$
\begin{equation}
\label{lmu}
	x_{n} = 2 \mu \exp \left(-1  -\frac{\pi (n - \frac{1}{4})}{\mu}\right)\,,
\end{equation}
 and $\mu \ll 1$
\begin{equation}
\label{smu}
	x_{n} = 2 \exp\left(-\frac{\pi n}{\mu} - \gamma\right)\,,
\end{equation}
with $\gamma$ the Euler–Mascheroni constant and $n = 1,2, \ldots$.

\begin{figure}[htb]
	\includegraphics[width=0.9\linewidth]{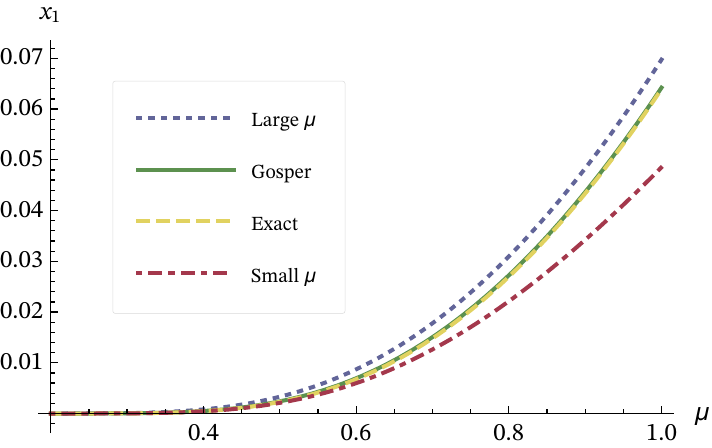}
		\caption{The first zero $x_1$ of $\mathcal{K}_{i \mu}(x)$ as a function of $\mu$.}
	\label{fig_app}
\end{figure}

Eq.~\eqref{smu} is widely used in the literature devoted to the $1/x^2$ potential~\cite{PhysRevD.48.5940,PhysRevLett.85.1590,PhysRevA.64.042103}. However, it is not relevant to our problem because vanishingly small SOI does not support the existence of the bound states. We are looking for the approximation valid from the intermediate $\mu \approx 0.5 \ldots 1$ to large values of $\mu$. Taking into account that in the intermediate case all $x_n \ll 1$, we expand the Macdonald function in the power series near $x = 0$~\cite{olver},
\begin{equation}
	\begin{split}
		\mathcal{K}_{i \mu}(x) = \frac{\pi}{2 \sin (i \pi \mu)} \bigg[&\frac{{(x/2)}^{-i \mu}}{\Gamma (1- i \mu) }\sum_{k=0}^{\infty} \frac{{(x^2/4)}^{k}}{k! {(1 - i\mu)}_{k}}\\ 
		{}- &\frac{{(x/2)}^{i \mu}}{\Gamma (1+ i \mu) }\sum_{k=0}^{\infty} \frac{{(x^2/4)}^{k}}{k! {(1 + i\mu)}_{k}}\bigg]\,.
	\end{split}
\end{equation}
Consequently, the zeros of $\mathcal{K}_{i \mu}(x)$ are determined from
\begin{equation}
	\begin{split}
		{\left(\frac{x}{2}\right)}^{2 i \mu} &= \frac{\Gamma (1+ i \mu)}{\Gamma (1- i \mu)} \dfrac{\sum \limits_{k=0}^{\infty} \dfrac{{(x^2/4)}^{k}}{k! {(1 - i\mu)}_{k}}}{\sum \limits_{k=0}^{\infty} \dfrac{{(x^2/4)}^{k}}{k! {(1 + i\mu)}_{k}}}\\
		&= -e^{2 i \arg \Gamma (i \mu)} + O(x^2)\,.
	\end{split}
\end{equation}
Taking the logarithm yields
\begin{equation}
	\log x_{n} = - \frac{\pi n}{\mu} + \log 2 + \frac{\pi}{2\mu} + \frac{1}{\mu} \arg \Gamma(i \mu)\,.
\end{equation}
We use the Gosper approximation~\cite{Gosper1978} for the Gamma-function to get
\begin{equation}
	\arg \Gamma (i \mu) \sim - \frac{\pi}{2} - \mu + \mu \log \mu + \frac{1}{2} \arctan (6\mu)\,.
\end{equation}
Finally, we obtain
\begin{equation}
\label{Gosper}
	x_{n}(\mu) = 2 \mu \exp \left(-1  -\frac{\pi n}{\mu} + \frac{1}{2 \mu} \arctan(6 \mu)\right)\,.
\end{equation}
For large $\mu$ this formula coincides with Eq.~\eqref{lmu}. The accuracy of this formula for intermediate values of $\mu$ can be checked by comparison with the exact result and approximations of Eq.~\eqref{lmu} and~\eqref{smu} in Fig.~\ref{fig_app}.

\bibliography{paper}

\end{document}